\documentstyle[12pt,fleqn]{article}
\topmargin - 0.8cm

\catcode`@=11
\@addtoreset{equation}{section}
\catcode`@=12
\mathchardef\SGamma="7100
\begin{document}
\title{\vspace{-0cm}\bf Decoherence in quantum cosmology at
the onset of inflation}
\author{A. O. Barvinsky$^{1}$, \
A. Yu. Kamenshchik$^{2,3}$, C. Kiefer$^{4}$ and \
 I. V. Mishakov$^{5}$}
\date{}
\maketitle
\hspace{-6mm}
$^{1}${\em Theory Department, Lebedev Institute and Lebedev
Research Center in Physics, Leninsky Prospect 53, Moscow 117924,
Russia}\\
$^{2}${\em L. D. Landau Institute for Theoretical Physics,
Russian Academy of Sciences, Kosygin Street 2, Moscow, 117334,
Russia}\\
$^{3}${\em
Landau Network-Centro Volta, Villa Olmo, Via Cantoni 1,
22100 Como, Italy}\\
$^{4}${\em Fakult\"at fur Physik, Universit\"at Freiburg,
 Hermann-Herder-Strasse 3, D-79104 Freiburg, Germany}\\
$^{5}${\em ``Polyprom M'', Komsomolsky Prospect 13, Moscow, Russia}\\

\begin{abstract}
We calculate the reduced density matrix for the inflaton field in a
model of chaotic inflation by tracing out degrees of freedom
corresponding to various bosonic fields. We find a qualitatively new
contribution to the density matrix given by the Euclidean effective
action of quantum fields. We regularise the ultraviolet divergences in
the decoherence factor. Dimensional regularisation
is shown to violate the consistency conditions for a density matrix as a
bounded operator. A physically motivated conformal redefinition of the
environmental fields leads to well-defined expressions. They show that
due to bosonic fields the Universe acquires classical properties near the
onset of inflation.
\end{abstract}

\newpage
\section{Introduction}
\hspace{\parindent}
The combination of general relativity with particle physics can yield
viable models for the early Universe. This leads, in particular, to the
idea that the Universe underwent a period of accelerated expansion
(``inflation") at an early stage. Inflation not only looses the dependence
on initial conditions, it also provides a quantitative scenario
for structure formation. The interesting question is of course why and
how inflation arises in the first place. It is generally assumed that
a satisfactory answer can only be found in the realm of quantum cosmology.
In this respect, ideas about the quantum nucleation of the Universe
from ``nothing" could yield a quantitative scenario for the
emergence of classical inflation. The problem of classical initial
conditions is then replaced by the question for an appropriate choice
of the quantum cosmological wave function.  Well-known choices
are the no-boundary proposal \cite{no-boundary} and the
tunneling proposal \cite{tun}, which are in fact a whole class of
proposals. There exist other interesting proposals such as the SIC
put forward in \cite{CZ}, that is in particular well equipped to deal
with the case of a classically recollapsing universe.

Inflation is assumed to take place near the GUT mass scale that is thought
to lie about five orders of magnitude below the Planck scale.
For this reason it is not necessary to rely on the most
fundamental level of quantum gravity (such as superstring theory),
and effective theories such as canonical quantum gravity should give
an excellent approximation. Using either the Euclidean path integral or
the Wheeler-DeWitt equation, the ``transition" from the classically
forbidden region to the classical inflationary regime has been
investigated in great detail. Using normalised wave functions,
a probability peak for the mass scale of inflation has been calculated
in various models \cite{qsi,qcr1}. Such a probability peak
has been interpreted -- assuming that we inhabit a generic Universe --
as providing a criterium to select amongst the members of an
``ensemble" of classical universes.

However, quantum theory does not yield a classical ensemble.
Since interference effects can play a crucial role, it would be
inconsistent to interpret these results directly as giving
probabilities for inflationary universes with different Hubble parameters.
A complete analysis would thus have to include a quantitative discussion
of the quantum-to-classical transition.

It is now generally accepted that the classical properties for a subsystem
arise from the irreversible interaction of this quantum system with its
natural environment. Starting with the pioneering work of Zeh
in the seventies \cite{Zeh}, this concept of {\em decoherence}
has been developed extensively, see \cite{decoherence1} and
\cite{decoherence2} for reviews. Quite recently, this continuous
loss of coherence was observed in quantum-optical experiments
\cite{Haroche}.

Quantum-to-classical transition through decoherence has also proven
fruitful in quantum cosmology. The first application of this concept
to gravitational systems was performed in a Newtonian framework
where the (formally quantised) gravitational field acquired
classical properties through interaction with masses \cite{Joos}.
In quantum cosmology, it was suggested to consider background degrees
of freedom (such as the scale factor) as the relevant system and
higher-order perturbations (such as gravitational waves or
density fluctuations) as an environment \cite{Zeh2}.
A quantitative discussion of this idea was first done in \cite{Kiefer1},
where it was demonstrated how the scale factor and the inflaton field
can acquire classical properties. This idea was further pursued
in many papers \cite{decoherence1}.
The quantum-to-classical transition through decoherence plays also
a crucial role for the primordial fluctuations
that eventually serve as classical seeds for galaxies and clusters of
galaxies \cite{Polarski}.

One big problem that remained in most of the above papers was the
issue of regularisation. Since there are infinitely many environmental
modes involved, there arise formal divergences that have to be dealt with.
This was done in most cases only heuristically through the choice of
a more or less appropriate cutoff in the number of modes.
It was argued in \cite{Paz} that dimensional regularisation is only
applicable for the phase part of the decoherence factor, not its
absolute value. This observation will be confirmed in detail in our
paper. Motivated by the finiteness of the decoherence factor in
QED, a redefinition of the fields was performed in a particular
quantum cosmological model in \cite{Claus} to find a finite result.
(A similar proposal was discussed recently in \cite{Okamura}.)
In the present paper we shall find in a wide class of models that there
exists a distinguished redefinition that renders the decoherence factor
finite.

A redefinition of environmental degrees of freedom changes, however,
the reduced density matrix for the system (in particular, the
system momentum is changed). This was discussed in \cite{LL}, where it was
suggested to use a redefinition that eliminates that part of the
 decoherence effect
that solely arises from the change in three-volume, so that only
decoherence due to ``particle creation" remains. It is thus clear that
a physical principle has to be invoked to select the appropriate
choice of environmental variables.

The purpose of our paper is to explore the possibilities to get a
physically reasonable, finite, decoherence factor that can
sensibly be calculated to understand the onset of inflation.
For this purpose, the non-diagonal elements in the reduced density
matrix have to be discussed. Using the criterium of normalisability
for the wave function in one-loop quantum cosmology, the diagonal
part was already discussed in the earlier papers
\cite{norm,qsi,tvsnb,qcr1,norm-napoli} where the focus was on the
derivation of probabilities. The same criterium will be employed here
for the study of decoherence. We shall apply our discussion to
a variety of models. The case of fermions, however, is discussed in
a separate paper \cite{fermions}, since a new formalism has to be applied
and novel features appear.

Our paper is organised as follows. In section~2 we briefly review
the form of the quantum cosmological wave function when the higher-order
modes are in their vacuum state. We then give a derivation of the correct
normalisation for this state, which is different from a (wrong)
normalisation that is sometimes used in the literature.
In section~3 we discuss the diagonal part of the reduced density matrix
for the inflaton field and, in particular, its dependence on the
one-loop effective action. Sections~4--6 comprise the main part of our paper.
In section~4 we discuss general properties of the non-diagonal elements
of the reduced density matrix. In section~5 we demonstrate that the
application of dimensional regularisation violates basic properties
of the density matrix. Section~6 then shows that there exists a
distinguished redefinition of the environmental degrees of freedom
that renders the decoherence factor finite. Section~7 is devoted
to a physical interpretation of the results obtained and an outlook
on future work.

\section{Cosmological wave function}
\hspace{\parindent}
We consider the quantum-cosmological wave function in the context
of chaotic inflation. This includes all cases of boundary conditions
mentioned in the Introduction \cite{no-boundary,tun,CZ}.
The background variables are the scale factor $a$ of the
inflationary universe and the inflaton field $\varphi$.
In addition we assume the presence of spatially inhomogeneous fields
$f({\bf x})$ that may be the higher modes of the inflaton or some other
field.
This fields play the role of microscopic (environmental) modes and are
treated perturbatively. They will thus be integrated out below to
study their decohering influence on the background variables.

 We shall consider the cosmological quantum state in the
framework of the reduced phase space quantisation, that is, in the
representation of physical variables. This has the advantage that a
well-defined inner product is available for the interpretation
of the density matrix. We must emphasise, though, that this
reduced quantisation is limited to semiclassical branches of the
wave function, but this is the framework where the present
investigation takes place.

In this representation the role of $f({\bf x})$ is played by physical
polarisations of linear fields of all  possible spins, and there is one
physical degree of freedom in the homogeneous sector of $(a,\varphi)$,
which without loss of generality can be identified with $\varphi$.
Thus, the full set of physical variables is
        \begin{equation}
        \phi(x) = [\,\varphi(t), f(t,{\bf x})\,],\,\,\,  \label{2.1}
        x=(t,{\bf x}).
        \end{equation}
With the decomposition of $f(t,{\bf x})$ into a discrete series of
spatial orthonormal harmonics $Q_n({\bf x})$ on a section of a
closed three-space,
        \begin{equation}
        f(t,{\bf x})=
        \sum_n f_n(t)Q_n({\bf x}),   \label{2.2}
        \end{equation}
$\phi(x)$ can equally be represented by the countable set of $(\varphi(t),f_n(t))$.

Other variables (including lapse and shift functions, unphysical components of
linear vector and tensor fields) are parametrised in terms of physical ones within
a particular choice of gauge conditions fixing the local gauge and
coordinate symmetries. In the sector of $(a,\varphi)$, for example, in the
cosmic-time gauge this parametrisation can have a very simple form
in the slow-roll approximation, corresponding to slowly varying (practically
constant) inflaton field. It can be taken to coincide
with an approximate classical solution with the cosmic time $t$,
        \begin{eqnarray}
        &&\varphi(t)=\varphi,     \\
        &&a(t) = \frac{1}{H} \cosh Ht,\,\,\, H=H(\varphi),   \label{a-Lorentz}
        \end{eqnarray}
where $H(\varphi)=8\pi V(\varphi)/3m_P^2$ is the Hubble constant
generated by the inflaton potential $V(\varphi)$. We emphasise that the
time parameter that appears in the reduced formalism is equivalent
to the standard WKB-time parameters that appears in the semiclassical
approximation to the Wheeler-DeWitt equation \cite{Kiefer2}.

In the no-boundary or tunneling prescription of the cosmological
wave function, the classical background (\ref{a-Lorentz}) arises
as an analytic continuation of the solution of Euclidean Einstein
equations,
        \begin{eqnarray}
        &&\varphi(\tau)=\varphi,     \nonumber\\
        &&a(\tau) = \frac{1}{H} \sin H\tau,\,\,\, H=H(\varphi), \label{2.3}
        \end{eqnarray}
into the complex plane of Euclidean time $\tau$,
        \begin{equation}
        \tau = \frac{\pi}{2 H}+i t\ .              \label{time}
        \end{equation}
This is usually interpreted as a
 ``quantum nucleation" of the Lorentzian DeSitter
spacetime from the gravitational instanton -- the Euclidean four-dimensional
hemisphere with radius $R=1/H(\varphi)$. In more standard language,
the Euclidean section of this instanton just corresponds to a classically
forbidden region, and the analytic continuation into the Lorentzian
regime corresponds to the emergence of time from timeless quantum
gravity \cite{Ringsberg}.
 The cosmological wave function
in Lorentzian time can also be understood as an analytic continuation of
the Euclidean wave function, which in the semiclassical one-loop
approximation has the form
        \begin{eqnarray}
        \Psi_{\rm E}(\tau|\varphi,f)=\frac1{\sqrt{u_\varphi(\tau)}}\,
        e^{-I(\tau,\varphi)}
        \prod_n \psi_n^{\rm E}(\tau,\varphi|f_n) \ .   \label{2.4}
        \end{eqnarray}
Here $I(\tau,\varphi)$ denotes the Euclidean Hamilton-Jacobi function for the
solution (\ref{2.3}), and
        \begin{eqnarray}
        \psi_n^{\rm E}(\tau,\varphi|f_n)=\frac1{\sqrt{u_n(\tau)}}
        \exp\left(-\frac12 a^3(\tau)\frac{\dot{u}_n(\tau)}
        {u_n(\tau)}f_n^2\right),                            \label{2.5}
        \end{eqnarray}
where $u_n(\tau)$ is the set of basis functions of the Euclidean linearised
equations of motion for the Euclidean action $I[\varphi,f]$ of physical
variables; it satisfies the regularity condition on the Euclidean
hemisphere containing the pole $\tau=0$,
        \begin{eqnarray}
        &&F_n(d/d\tau)u_n(\tau)=0,\,\,\,u_n(\tau)={\rm reg},  \label{2.6}\\
        &&F_n(d/d\tau)\delta(\tau-\tau') =
        \frac{\delta^{2}I[\varphi,f]}{\delta f_n(\tau)
        \delta f_n(\tau')}\ .                              \label{operator}
        \end{eqnarray}
Analogously, $u_\varphi(\tau)$ in (\ref{2.4}) denotes the basis function of
a spatially homogeneous inflaton mode.
This expression was obtained both by solving the Wheeler-DeWitt equation
and by calculating the Euclidean path integral
 \cite{tvsnb,Laf,VilVach,tunnel}\footnote{Here, for simplicity,
 we consider the case of a scalar field with
$F_n(d/d\tau)=-(d/d\tau)a^3(d/d\tau)+...$, while for the generic case
\cite{tunnel} $a^3\dot{u}_n$ in the exponential should be replaced by the
corresponding Wronskian operator linear in time derivative acting on $u_n$.
}.
It is important that in the above expressions the basis functions and
Gaussian states of quantum variables $f_n$ are all calculated on the background
of the classical solution parametrised by the inflaton variable
 $\varphi$ which,
as an argument of the wave function, is also an essential quantum degree
of freedom.

The analytic continuation to the Lorentzian wave function
        \begin{equation}
        \Psi(t|\varphi,f)=\Psi_{\rm E}(\pi/2H+it|\varphi,f)  \label{2.7}
        \end{equation}
takes place by introducing the time-independent Euclidean action $I(\varphi)$
of the DeSitter instanton and the Lorentzian Hamilton-Jacobi function
$S(t,\varphi)$,
        \begin{eqnarray}
        &&I(\pi/2H+it,\varphi)=\frac12
        I(\varphi)-iS(t,\varphi),                           \label{2.8}\\
        &&I(\varphi)=2I(\pi/2H,\varphi)\simeq
        -\frac{3m_P^4}{8V(\varphi)}\ ,                         \label{2.9}
        \end{eqnarray}
and the basis functions of the Lorentzian linearised field equations,
        \begin{eqnarray}
        &&v_n(t) = [u_n(\pi/2H+it)]^{*},\,\,\,
        v^{*}_n(t) = u_n(\pi/2H+it)],            \label{modes-cont}\\
        &&F^{\rm L}_n(d/dt) v_n(t) = 0,\,\,\,
        F^{\rm L}_n(d/dt)\delta(t-t')=
        \frac{\delta^{2}S}{\delta f_n(t) \delta f_n(t')}\ .  \label{2.10}
        \end{eqnarray}
The latter turn out to be the basis functions of the DeSitter-invariant
Bunch-Davies vacuum \cite{Laf,Allen}. The cosmological wave function thus
takes the form of the product
        \begin{eqnarray}
        \Psi(t|\varphi,f)=\frac1{\sqrt{v^*_\varphi(t)}}\,
        e^{-I(\varphi)/2+iS(t,\varphi)}
        \prod_n \psi_n(t,\varphi|f_n)                     \label{2.11}
        \end{eqnarray}
of the background wave function and the infinite set of Gaussian states of
harmonic oscillators representing this vacuum,
        \begin{eqnarray}
        &&\psi_n(t,\varphi|f_n)=\frac1{\sqrt{v^*_n(t)}}
        \exp\left(-\frac12 \Omega_n(t) f_n^2\right),       \label{2.12}\\
        &&\Omega_n(t)=-ia^3(t)\frac{\dot{v}^*_n(t)}{v^*_n(t)}.  \label{2.13}
        \end{eqnarray}
An answer similar to (\ref{2.11}) holds also for the case of the
tunneling wave function -- the only difference is the opposite sign of the
Euclidean action in the tree-level part. In what follows we shall mainly
consider the no-boundary wavefunction, but the main conclusions and
algorithms for the decoherence aspects of the system will be equally
applicable in the tunneling case as well\footnote
{In the case of the tunneling wave function, the Euclidean path integral
derivation does not work, and it can be obtained as an alternative
solution of the Wheeler-DeWitt equation linearly independent from the
no-boundary one. Strictly speaking, the behaviours of both of these
wave functions in the Euclidean and Lorentzian domains does not reduce to a
simple analytic continuation presented above. For a no-boundary state one
underbarrier branch matches with two complex-conjugated branches in the
Lorentzian domain, while for the tunneling state the situation is
reversed. For our presentation here these subtleties are immaterial. We
shall briefly get back to this issue in the concluding section when
discussing the decoherence between the branches.
}.

An important property of these vacuum states is that their norm
is conserved along any semiclassical solution (\ref{a-Lorentz}),
        \begin{eqnarray}
        &&\Big<\psi_n,\psi_n\Big>\equiv\int
        df_n|\psi_n(f_n)|^2=
        \sqrt{2\pi} [\Delta_n(\varphi)]^{-1/2},      \label{2.14}\\
        &&\Delta_n(\varphi)\equiv
        ia^3(v^*_n\dot{v}_n-\dot{v}^*_n v_n)=\mbox{constant}\ .   \label{2.15}
        \end{eqnarray}
Note that $\Delta_n(\varphi)$ is just the (constant) Wronskian
corresponding to (\ref{2.10}). We want to emphasise, however,
that $\Delta_n$ is
a nontrivial function of the background variable $\varphi$, since it is
defined on full configuration space and not only along a semiclassical
trajectory.
It can thus not be factored out by a deliberate normalisation of these states
to unity -- a procedure that is sometimes assumed in
the literature \cite{Okamura}.
The replacement of $\psi_n(t,\varphi|f_n)$ with the normalised states
        \begin{eqnarray}
        &&\phi_n(t,\varphi|f_n)=
        \left(\frac{\Delta_n}{2\pi}\right)^{1/4}
        \psi_n(t,\varphi|f_n)
        =\left(\frac{{\rm Re}\,\Omega_n}{\pi}\right)^{1/4}
        \left(\frac{v_n}{v_v^*}\right)^{1/2}
        \exp\left(-\frac12 \Omega_n f_n^2\right),        \label{2.16}\\
        &&\Big<\phi_n,\phi_n\Big>=1
        \end{eqnarray}
would be inconsistent from the viewpoint of the full Wheeler-DeWitt equation
or the path-integral representation of its solution, because the multiplication
of this solution by a configuration space-dependent object would violate the
Wheeler-DeWitt equation. We shall give in the rest of this section
the justification of the use of the unnormalised vacuum
wave functions in (\ref{2.12}).

\subsection{Justification of wave function prefactors: reduction method for
functional determinants}
\hspace{\parindent}
One type of justification comes from the general semiclassical theory of
constrained dynamics \cite{BK} including the Wheeler-DeWitt equation as a
particular case. As shown in \cite{BK}, there exists a unitary map between
the solutions of the quantum Dirac constraints and physical wave functions
in reduced phase-space quantisation. These wave functions, like
(\ref{2.5}),
have the form of a semiclassical packet with a prefactor given by the
Jacobian of the transformation from the initial coordinates of the flow of
classical trajectories to their final coordinates. The matrix of this
Jacobian is the solution of linearised equations of motion, just like the
basis functions in the prefactors of (\ref{2.5}), and these basis
functions are not normalised to unity in the sense of a Wronskian inner product
(\ref{2.15}).

 A more useful justification of this fact comes from
the Euclidean path-integral representation of the no-boundary
wave function, which goes as follows.
The Hartle-Hawking path integral in the one-loop approximation reads
        \begin{eqnarray}
        \Psi_{\rm E}^{\rm 1-loop}(\tau|\varphi,f)=
        \left.\int D\phi(x)e^{-I[g(x)]}\right\vert_{\rm one-loop}
        =e^{-I(\tau|\varphi,f)}\,
        \left[\,{\rm Det} F\,\right]^{-1/2}\ .           \label{2.17}
        \end{eqnarray}
It contains the tree-level classical action which when expanded up to
quadratic order in microscopic variables,
        \begin{eqnarray}
        I(\tau|\varphi,f)=I(\tau,\varphi)+
        \frac12 a^3(\tau)\frac{\dot{u}_n(\tau)}
        {u_n(\tau)}f_n^2\ ,                          \label{2.18}
        \end{eqnarray}
gives rise to exponentials of Gaussian oscillator functions; it also contains
the one-loop functional determinant of the operator
        \begin{equation}
        F = \frac{\delta^{2}I[\phi]}
        {\delta\phi(x)\delta\phi(y)}      \label{2.19}
        \end{equation}
subject to Dirichlet conditions at the boundary of the Euclidean spacetime
ball of ``radius'' $\tau$. These boundary conditions follow from the fact
that the integration in the path integral requires fixed values of
fields at the boundary as prescribed by the arguments of the wave function.
The definition of the functional determinant can then be given in terms of
the eigenvalues of the corresponding eigenvalue problem
        \begin{eqnarray}
        &&F\phi_\alpha(\tau')=\lambda_\alpha\phi_\alpha(\tau'),
        \,\,\,0\leq\tau'\leq\tau,
        \,\,\,\phi_\alpha(\tau)=0,\,\,
        \lambda_\alpha >0,                       \label{2.20}\\
        &&\ln\,{\rm Det}\,F={\rm Tr}\,
        \ln\,F=\sum_\alpha
        \ln\lambda_\alpha.                        \label{2.21}
        \end{eqnarray}
Here $\alpha$ is a collective index of arbitrary nature enumerating these
eigenvalues.

Suppose now that we can write down an equation which
determines these eigenvalues in the form
        \begin{eqnarray}
        E(\lambda)=0.                               \label{2.22}
        \end{eqnarray}
Its left-hand side $E(\lambda)$ can be used to represent the sum over the
eigenvalues as an integral in the complex plane of $\lambda$ over the
contour $C_+$ surrounding the spectrum of the operator -- so that the
residues in the roots of (\ref{2.22}) reproduce the contributions
of different eigenvalues
        \begin{eqnarray}
        \sum_\alpha \ln\lambda_\alpha=\frac1{2\pi i}\int\limits_{C_+}
        d\lambda\,\ln\lambda\,\frac d{d\lambda}\ln E(\lambda).        \label{2.23}
        \end{eqnarray}
In what follows we assume that this spectrum is positive and lies on the
positive real axis. Omitting some technical details \cite{annals} we integrate
in this integral by parts and rotate the contour of integration into
 the contour $C_-$ so that it begins
surrounding the origin and running over the upper and lower shores of the
negative real axis. In view of the absence of other roots in the complex
plane of $\lambda$, the only arising residue of the resulting integral
gives the final answer in terms of the left-hand side of the eigenvalue
equation
        \begin{eqnarray}
        \sum_\alpha \ln\lambda_\alpha=
        -\frac1{2\pi i}\int\limits_{C_-}
        \frac{d\lambda}{\lambda}\,\ln E(\lambda)=\ln E(0).  \label{2.24}
        \end{eqnarray}

In the basis of orthonormal spatial harmonics the operator has a diagonal
form, so that the total equation on its eigenvalues decomposes into an
infinite product of partial equations
        \begin{eqnarray}
        F={\rm diag}\{F_n(d/d\tau)\},\,\,\,
        E(\lambda)=\prod_n E_n(\lambda),                     \label{2.25}
        \end{eqnarray}
and the final answer reads
        \begin{eqnarray}
        \sum_\alpha \ln\lambda_\alpha=\sum_n\ln E_n(0)\equiv
        \sum_{n=0}^{\infty}
        {\rm dim}(n)\,\ln E_n(0),                             \label{2.26}
        \end{eqnarray}
where the summation over the collective index of these harmonics is
rewritten as a sum involving the degeneracies of the corresponding quantum
numbers.

Let us now get back to the calculation of a particular functional
determinant in the no-boundary wave function. The role of eigenfunctions in
this case is played by the basis functions of the operator modified by the
extra (mass) term, satisfying the condition of regularity at $\tau=0$,
        \begin{eqnarray}
        (F_n(d/d\tau')-\lambda)u_{n,\lambda}(\tau')=0,
        \,\,\,0<\tau'<\tau.                                 \label{2.27}
        \end{eqnarray}
Therefore the left-hand side of the eigenvalue equation takes the form
        \begin{eqnarray}
        E_n(\lambda)=u_{n,\lambda}(\tau),                     \label{2.28}
        \end{eqnarray}
and its value at zero argument coincides with the original regular basis
function $E_n(0)=u_{n,0}(\tau)=u_n(\tau)$ taken at $\tau$ -- the ``radial''
coordinate of the boundary. The use of the equation (\ref{2.26}) then
immediately gives
        \begin{eqnarray}
        \left[{\rm Det} F\right]^{-1/2}=\exp\left(-\frac12\sum_\alpha
        \ln\lambda_\alpha\right)=
        \frac1{\sqrt{u_\varphi(\tau)}}\prod_n
        \frac1{\sqrt{u_n(\tau)}}\, ,                              \label{2.29}
        \end{eqnarray}
where we have also included the contribution of the inflaton sector of
the total diagonal operator
$F={\rm diag}\{F_\varphi(d/d\tau),F_n(d/d\tau)\}$.

The above derivation is presented in a simplified form, because it
disregards special measures to provide the vanishing of the arc integrals at
the infinity of the complex plane in the transition from (\ref{2.23}) to
(\ref{2.24}), regularisation of ultraviolet divergences and the
precautions against spurious roots of (\ref{2.26}) caused by
multiplication of $E_n(\lambda)$ by functions of $\lambda$. These
problems are considered in much detail in the authors' paper \cite{annals},
where it is shown, in particular, that the normalization of the Euclidean basis
functions in (\ref{2.28}) and (\ref{2.29}) should be chosen in such a way
that the coefficient of their power-like asymptotic behaviour at $\tau=0$,
        \begin{eqnarray}
        u_n(\tau)\simeq U_n\tau^{\mu_n},\,\,\,\tau\rightarrow 0,
        \end{eqnarray}
is a trivial field-independent constant. Such a normalisation is very
different from the normalisation to unity with respect to the Wronskian
inner product.

\section{Wave function norm, Euclidean effective action, and the
diagonal of the density matrix}
\hspace{\parindent}
The nontrivial normalisation of vacuum states results in a nontrivial
quantum distribution of cosmological parameters in inflaton space,
yielding a probability distribution for the inflaton. This
mechanism was used in \cite{norm,BarvU} to analyse
 the normalisability of the
cosmological wave function and to find initial conditions for inflation
\cite{qcr1,qsi,tvsnb}. We briefly repeat these results here, for they
will play a substantial role also in the decoherence properties of the
density matrix.

Consider the quantum distribution function of the inflaton field which is
just the diagonal element of the reduced density matrix obtained by
tracing out the microscopic variables,
        \begin{eqnarray}
        \rho(t|\varphi)\equiv\rho(t|\varphi,\varphi) =
        \int df\,|\Psi(t|\varphi,f)|^2.                     \label{3.1}
        \end{eqnarray}
It involves the infinite product of norms (\ref{2.14})-(\ref{2.15}) that
can be reinterpreted by using the reduction technique for functional
determinants of the previous section. For this, let us introduce two sets
of Euclidean basis functions inhabiting the total spherical manifold of the
DeSitter-instanton (the full sphere with the scale factor (\ref{2.3})
extended to the full range of the latitude angle $0<\tau/H<\pi$),
        \begin{eqnarray}
        F^{\rm E}(d/d\tau)u^\pm(\tau)=0,\,\,\,
        u^-(\tau)=u(\tau),\,\,\,u^+(\tau)=u(\pi/H-\tau).  \label{3.2}
        \end{eqnarray}
They are related by mirror map with respect to the equatorial section of
the instanton $\tau=\pi/2H$ at which the nucleation into the Lorentzian domain
takes place. Generically, $u_-(\tau)$ is regular at $\tau=0$ but singular
at the antipodial pole of the sphere, $\tau=\pi/H$, and vice versa for
$u_+(\tau)$. In the Lorentzian domain they give rise to negative and positive
frequency basis functions
        \begin{eqnarray}
        u^-(\pi/2H+it)=v^*(t),\,\,\,u^+(\pi/2H+it)=v(t),   \label{3.3}
        \end{eqnarray}
the Wronskian norm of which can thus be rewritten as a Wronskian product
of two different Euclidean basis functions calculated at $\tau=\pi/2H$,
        \begin{eqnarray}
        \Delta_n(\varphi)\equiv ia^3\Big(v^*_n\dot{v}_n-\dot{v}^*_n
        v_n\Big)_{t=0}=a^3\Big(u^+\frac{du^-}{d\tau}
        -\frac{du^+}{d\tau}u^-\Big)_{\tau=\pi/2H}\ .             \label{3.4}
        \end{eqnarray}

Let us now modify these functions by the $\lambda$-term in analogy with
the previous section, $u_\lambda^{\pm}(\tau)$,
        \begin{eqnarray}
        (F_n(d/d\tau)-\lambda)u_\lambda^{\pm}(\tau)=0,\,\,\,
        0<\tau<\pi/H,                                          \label{3.5}
        \end{eqnarray}
still demanding the regularity of $u_\lambda^-(\tau)$ at $\tau=0$ and of
$u_\lambda^+(\tau)$ at $\tau=\pi/H$, and define the function
        \begin{eqnarray}
        E_n(\lambda)=a^3\Big(u^+_\lambda
        \frac{du^-_\lambda}{d\tau}
        -\frac{du^+_\lambda}{d\tau}u^-_\lambda\Big)_n.     \label{3.6}
        \end{eqnarray}
Equation (\ref{2.28}) with this function will be just the requirement
of the {\em functional} linear dependence of $u_\lambda^-(\tau)$ and
$u_\lambda^+(\tau)$ (in view of the second-order equation of motion the
equality of the functions and their derivatives at one point implies their
equality at any $\tau$). This linear dependence means that there are
solutions of (\ref{3.5}) that are regular on the {\em whole} instanton
including both its poles at particular values of $\lambda$ satisfying this
equation. These values represent the spectrum of the operator on the {\em whole}
DeSitter instanton, and according to (\ref{2.26}) the product
of norms arising in (\ref{3.1}) reads, up to a numerical constant, in terms
of this spectrum,
        \begin{eqnarray}
        \prod_n\Big<\psi_n,\psi_n\Big>={\rm C}\sqrt{\Delta_\varphi}
        \exp\left(-\frac12\sum_\alpha
        \ln\lambda_\alpha\right)={\rm C}
        \sqrt{\Delta_\varphi}
        \exp\left(-\frac12{\rm Tr}\,\ln F\right)\ .           \label{3.7}
        \end{eqnarray}
It can thus be expressed
as the one-loop Euclidean effective action of the theory calculated on the
DeSitter-instanton of the radius $1/H(\varphi)$
 \cite{norm,tvsnb,tunnel,BarvU},
        \begin{eqnarray}
        \mbox{\boldmath$\SGamma$}_{\rm 1-loop}(\varphi)
        =\frac12{\rm Tr}\,\ln F \ .                      \label{3.8}
        \end{eqnarray}
(The total one-loop action contains the contribution of the
 inflaton mode missing in this equation;
this explains the origin of an additional $\Delta_\varphi$ factor
 in (\ref{3.7}).)

Thus, the distribution function takes the form
        \begin{eqnarray}
        \rho(t|\varphi,\varphi)={\rm C}
        \frac{\sqrt{\Delta_\varphi}}{|v_\varphi(t)|}
        \exp(-I(\varphi)-
        \mbox{\boldmath$\SGamma$}_{\rm 1-loop}(\varphi)).        \label{distr}
        \end{eqnarray}
The derivation of this result was presented here for the case of the
no-boundary wave function. It is valid, however, also in the tunneling case
of the model with microscopic oscillator modes \cite{VilVach}. The basic
difference in this case is the opposite sign of the tree-level Euclidean
action $I(\varphi)$ in the exponential.

The advantage of this representation is that there exist powerful methods
of calculation and covariant renormalisation of
$\mbox{\boldmath$\SGamma$}_{\rm 1-loop}(\varphi)$. These methods, in
particular, lead to the asymptotic scaling behaviour
        \begin{eqnarray}
        \mbox{\boldmath$\SGamma$}_{\rm 1-loop}(\varphi)
        =Z\ln\frac{H(\varphi)}{\mu},\,\,\,
        H(\varphi)\rightarrow\infty,                          \label{effect}
        \end{eqnarray}
with an easily calculable coefficient $Z$ \cite{qsi,qcr1}. For big positive
$Z$ the one-loop contribution guarantees the normalisability of the
cosmological wave function at large $\varphi$. In the model with large
negative nonminimal coupling of the inflaton it leads to a sharp
probability peak with the energy-scale parameters belonging to the GUT
domain rather than the Planck domain \cite{qsi,qcr1} and serves as a
``source" for inflation.

\section{The density matrix}
\hspace{\parindent}
We now proceed to discuss the off-diagonal elements of the reduced
density matrix that are needed for the discussion of decoherence.
We shall restrict ourselves to one semiclassical branch, i.e.
to a wave function of the form (\ref{2.11}). We shall comment
on decoherence {\em between} branches in section~7.

The off-diagonal elements of the reduced density matrix read
        \begin{equation}
        \rho(t|\varphi,\varphi') = \int df
        \Psi(t|\varphi,f)\Psi(t|\varphi',f)\ ,           \label{density}
        \end{equation}
where $\Psi$ is given by (\ref{2.12}).
After the Gaussian integration one obtains
        \begin{eqnarray}
        \rho(t|\varphi,\varphi')=
        {\rm C}\frac{\Delta_\varphi\Delta_\varphi'}
        {\sqrt{v_\varphi(t) v'_\varphi(t)}}
        \exp\Big[-\frac12I-\frac12I'+i(S-S')\Big]\prod_n
        \Big[v_n^*v_n'(\Omega_n+\Omega_n'^*)\Big]^{-1/2}.     \label{4.1}
        \end{eqnarray}
It is useful to extract from this density matrix the factor
$\sqrt{\rho(t|\varphi,\varphi)\rho(t|\varphi',\varphi')}$,
cf. (3.10), the remaining factor
describing the dynamical entanglement during the Lorentzian evolution. By using
the relation $\Delta_n=2|v_n|^2{\rm Re}\,\Omega_n$ and the expression for
the one-loop effective action obtained above, we get
        \begin{eqnarray}
        &&\rho(t|\varphi,\varphi')=
        {\rm C}\frac{\Delta_\varphi\Delta_\varphi'}
        {\sqrt{v_\varphi(t) v'_\varphi(t)}}
        \exp\left(-\frac12\mbox{\boldmath$\SGamma$}
        -\frac12\mbox{\boldmath$\SGamma$}'
        +i(S-S')\right)\mbox{\boldmath$D$}(t|\varphi,\varphi'),  \label{4.2}
        \end{eqnarray}
where
        \begin{eqnarray}
        \mbox{\boldmath$\SGamma$}=
        I(\varphi)+
        \mbox{\boldmath$\SGamma$}_{\rm 1-loop}(\varphi)       \label{4.3}
        \end{eqnarray}
is the full Euclidean effective action including the classical part, and
        \begin{eqnarray}
        \mbox{\boldmath$D$}(t|\varphi,\varphi')=
        \prod_n\left(\,\frac{4{\rm Re}\,\Omega_n\,{\rm Re}\,\Omega_n'^*}
        {(\Omega_n+\Omega_n'^*)^2}\,\right)^{1/4}
        \left(\frac{v_n}{v_n^*}
        \frac{v_n^{'*}}{v_n'}\right)^{1/4}                   \label{4.4}
        \end{eqnarray}
is the conventional decoherence factor. The form (4.3)
of the density matrix corresponds to an ``ideal measurement" by the
environmental variables, since the diagonal part (the probabilities)
remains unchanged; only the coherence is affected.

It should be emphasised here that,
although the decoherence factor coincides with the one known in the
literature \cite{Paz,Claus,Okamura}, the {\em total}
density matrix (\ref{4.2})
is different, because it contains the effective-action factors related to
both arguments of $\rho(t|\varphi,\varphi')$. This is the effect of a
correct normalisation of the vacuum Gaussian states for microscopic
variables. The factors are entirely defined on the DeSitter-instanton and,
therefore, independent of the Lorentzian time. Therefore they play an
essential role only at the onset of inflation, provided the decoherence
effects rapidly grow during the Lorentzian evolution. As we shall see,
this will indeed be the case for all fields except for the case of massless
conformally invariant fields.

The amplitude of the decoherence factor can be rewritten in
the form
        \begin{eqnarray}
        |\mbox{\boldmath$D$}(t|\varphi,\varphi')|=
        \exp\frac14\sum_n\ln\frac{4{\rm Re}\,\Omega_n\,{\rm Re}\,\Omega_n'^*}
        {(\Omega_n+\Omega_n'^*)^2},            \label{4.5}
        \end{eqnarray}
and the convergence of this series is far from
being guaranteed \cite{Paz,Claus,Okamura}. Moreover, the divergences might
be not renormalisable by local counterterms in the bare quantised action.
We shall now analyse this question of coherence for all types of bosonic fields;
the case of fermions will be dealt with in a separate paper \cite{fermions}.

\section{Inconsistencies in the dimensionally regularised density matrix}
\hspace{\parindent}
In this section we shall show that dimensional regularisation
leads to density matrices that violate crucial properties
and that are therefore inconsistent. We shall start with the case
of a massive minimally coupled scalar field. There the
 equation for the Lorentzian basis functions reads
        \begin{eqnarray}
        \frac d{dt}\left(a^3\frac{dv_n}{dt}\right)
        +\left(\frac{n^2}{a^2}+m^2\right)v_n=0.           \label{5.1}
        \end{eqnarray}
In what follows we shall use dimensional regularisation to regulate the
divergencies. In this regularisation the equation retains its form except
for the power of $a$ in the first term, which instead of 3 becomes $d-1$ where
$d\rightarrow 4$ is the regularisation parameter of the spacetime dimension
(and $n$ is being changed to $n+(d-4)/2$).
The corresponding solutions for the DeSitter background, with the Hubble
constant $H$, that are regular on the Euclidean hemisphere of the
gravitational instanton (at $\tau=\pi/2H+it=0$) read in this dimension
\cite{annals}
        \begin{equation}
        v_n(t) = (\cosh Ht)^{-\frac{(d-2)}{2}}
        P^{-(n+\frac{d-4}{2})}_{-\frac{1}{2}
        +i\sqrt{m^2/H^2-(d-1)^2/4}}(i\sinh Ht).            \label{basis1}
        \end{equation}

The behaviour of the series (\ref{4.5}) depends on the behaviour of $\Omega_n$
with respect to $n$. Since we shall be strongly interested in the limit of
large masses, it is worth presenting the above function in the form of an
$1/m$ asymptotic expansion uniform in $n$ \cite{annals,asymp}. It reads
        \begin{eqnarray}
        &&P_{-1/2+i\lambda}^{-\mu}(z) =
        \frac{1}{|\Gamma(n+i\lambda+1/2)|}
        \frac{(\beta^{2}-z^{2})^{-1/4}}{\sqrt{\lambda}}      \nonumber \\
        &&\qquad\qquad\qquad\times
        \exp\left(\frac{\mu}{2}\ln\frac{1-v}{1+v}
        -\lambda\arctan\frac{v}{\alpha}\right)\,\sum_{k=0}^{\infty}
        \frac{T_{k}(v,\alpha)}{\lambda^{k}},\;\;
        \lambda\rightarrow \infty,                         \label{uniform}
        \end{eqnarray}
where
        \begin{eqnarray}
        \alpha = \frac\mu\lambda,\,\,
        \beta^{2} = 1 + \alpha^{2},\,\,
        v = \frac{\alpha z}{(\beta^{2} - z^{2})^{1/2}},
        \end{eqnarray}
and the coefficients $T_{k}(v,\alpha)$ beginning with $T_0(v,\alpha)=1$
are given in \cite{annals} (they are not needed below).
They are uniformly bounded for all values of $\lambda$ and $\mu$. In view of
this boundedness, the exponential part of this function
 represents for large $n$
the conventional high-frequency adiabatic expansion and contributes the dominant
part to $\Omega_n$,
        \begin{eqnarray}
        \Omega_n=a^2\left[\sqrt{n^2+m^2a^2}
        +i\sinh Ht\left(1+\frac12\frac{m^2a^2}
        {n^2+m^2a^2}\right)\right]+O(1/m).                \label{5.2om}
        \end{eqnarray}
The corresponding leading contribution to the amplitude of the
decoherence factor,
        \begin{eqnarray}
        \ln|\mbox{\boldmath$D$}(t|\varphi,\varphi')|\simeq\frac14
        \sum_{n=0}^\infty n^2\ln\frac{4a^2a'^2\sqrt{n^2+m^2a^2}\,
        \sqrt{n^2+m^2a'^2}}{\left(a^2\sqrt{n^2+m^2a^2}
        +a'^2\sqrt{n^2+m^2a'^2}\right)^2}\ ,              \label{5.3}
        \end{eqnarray}
obviously contains cubic and linear divergences which cannot be
represented as additive functions of $a$ and $a'$. This means that no
one-argument counterterm to $\mbox{\boldmath$\SGamma$}$ and
$\mbox{\boldmath$\SGamma$}'$ in (\ref{4.2}) can cancel these
divergences of the amplitude -- an observation made a number of years ago
in \cite{Paz,Claus}.

Let us forget for the moment this difficulty and just use the
regularisation which identically discards all power divergences. Then,
provided that the logarithmic divergences are vanishing, one can at least
have a finite quantity subject to physical analysis. Another problem can,
however, occur in this case -- this regularisation can perform an
``oversubtraction" in the sense
that the resulting regularised density matrix will be
inconsistent -- it will violate the necessary properties satisfied for a
convergent series (\ref{4.5}). The calculations briefly presented below
really show that the dimensionally-regularised density matrix would have
off-diagonal terms infinitely growing for
$\varphi-\varphi'\rightarrow\infty$ and thus being inconsistent;
for example, ${\rm Tr}\,\hat{\rho}^2$ would diverge. We emphasise that
this problem has not been discussed before, since reduced
density matrices are usually not considered in quantum field theory.

Dimensional regularisation of the divergent sum (\ref{5.3}) implies that
it should be reformulated in $d$-dimensional spacetime, which boils
down to two main modifications -- the degeneracy of the $n$-th eigenvalue of the
spatial Laplacian $n^2$, and the eigenvalues $\lambda_n=-n^2/a^2$ itself should be
replaced by their $d$-dimensional versions ${\rm dim}(n,d)$ and
$\lambda_n(d)$. From the theory of irreducible representations of the group
$O(d+1)$ it is well known that
        \begin{eqnarray}
        &&{\rm dim}(n,d) = \frac{(2n + d - 4)
        \Gamma(n + d - 3)}
        {\Gamma(n)\Gamma(d - 1)}= \nonumber\\
        &&\qquad\qquad\qquad n^{2-\epsilon}\,
        \frac{(2-\epsilon/n)(1-\epsilon/n)}{\Gamma(3-\epsilon)}
        \frac{\Gamma(n-\epsilon)\,n^\epsilon}{\Gamma(n)}
        \sim n^{2-\epsilon},\,\,\,\,\epsilon\equiv 4-d,    \label{scal-dimen}
        \end{eqnarray}
while the $d$-dimensional $\lambda_n(d)$ can be obtained from $\lambda_n$ by
shifting the quantum number $n$, $n\to n-(4-d)/2$. The latter results in
the following modification of the frequency,
        \begin{eqnarray}
        \Omega_n(d)=a^2\sqrt{\left(n-\frac{4-d}2\right)^2
        +m^2a^2}+O(m^0).                                         \label{5.2}
        \end{eqnarray}
Note that the main effect of regularisation in (\ref{scal-dimen}) consists in
the multiplication of the summation measure $n^2$ by the factor $n^{-\epsilon}$
which for large positive $\epsilon\equiv 4-d$ can make the divergent
series convergent. The analytic continuation of the result to $\epsilon=0$
allows one to disentangle the divergent part of the whole answer as a
pole in dimensionality and get a finite contribution.

The further calculations are based on the use of the summation
Euler-Maclaurin formula,
        \begin{eqnarray}
        \sum_{n=1}^\infty f(n)=\int_0^\infty dy f(y)-\frac12 f(0)
        +i\int_0^\infty dy \frac{f(iy)-f(-iy)}{e^{2\pi y}-1},
        \end{eqnarray}
which allows one to reduce the infinite sum to the combination of
integrals along real and imaginary axes in the complex plane of $n=y$.
Because of the exponential damping factor in the integral over imaginary
axes, the regularisation can be removed in its integrand from the very
beginning. This would mean that its contribution is vanishing (because the
integrand is an even function of its argument), except for
the possible residues
at the singular points of the integrand given by the equation
$n^2+(ma)^2\equiv -y^2+(ma)^2=0$. These residues will, however, contribute
the exponentially small terms $O(e^{-ma})$ negligible in the approximation
of large masses $ma\gg 1$. Thus, the decoherence factor reduces in the
main to the contribution of the integral over the real axes, which after
the change of the integration variable $y=ma/x$, takes the following form,
        \begin{eqnarray}
        \ln|\mbox{\boldmath$D$}|
        =\int_0^\infty dx\,x^{\epsilon-4}
        f(x,\epsilon)+O(e^{-ma}).                \label{5.5}
        \end{eqnarray}
Its integrand can be obtained from (\ref{5.3}) by the substitution of
$n=ma/x$ and the dimensional modifications of the above type, and for
nonvanishing $\epsilon$ it looks rather complicated. However, its
complexity due to the nontrivial dimensionality can be represented in the
following form,
        \begin{eqnarray}
        &&f(x,\epsilon)=\frac{2(ma)^{-\epsilon}}{\Gamma(3-\epsilon)}
        f(x,0)\left[\,1+O(\epsilon/ma)\,\right],     \label{5.6}\\
        &&f(x,0)=\frac14 (ma)^3\ln
        \frac{4\delta^2\sqrt{1+x^2}\,
        \sqrt{1+\delta^2x^2}}{\left(\sqrt{1+x^2}
        +\delta^2\sqrt{1+\delta^2x^2}\right)^2},\,\,\,      \label{5.7}
        \delta\equiv\frac{a'}a,
        \end{eqnarray}
with the function $f(x,0)$ trivially obtained from (\ref{5.3}) by
introducing the new parameter $\delta$. Due to this, the resulting
expression does not look symmetric in $a$ and $a'$, but it will allow us
easily to analyse the behaviour of far off-diagonal elements
corresponding to $\delta\ll 1$.

The further transformations of (\ref{5.5}) consist in the consecutive
integrations by parts in the convergence domain of the integral and the
analytic continuation to $\epsilon=0$:
        \begin{eqnarray}
        &&\int_0^\infty dx\,x^{\epsilon-4} f(x,\epsilon)=
        \frac{f'''(0,0)}{\epsilon}-\frac16\int_0^\infty dx\,\ln
        x\,f^{\rm (iv)}(x,0)\nonumber\\
        &&\qquad\qquad\qquad\qquad\qquad
         - \left.\frac{d}{d\epsilon}\frac{f'''(0,\epsilon)}
        {(\epsilon-3)(\epsilon-2)(\epsilon-1)}\right|_{\epsilon=0}.
        \label{5.8}
        \end{eqnarray}
In view of (\ref{5.6})-(\ref{5.7}) one has
        \begin{eqnarray}
        f'''(0,0)=0,\,\,\,\left.\frac{d}{d\epsilon}f'''(0,\epsilon)
        \right|_{\epsilon=0}=(ma)^3\,O(1/m),
        \end{eqnarray}
which means the absence of logarithmic divergences, and the final result
reads
        \begin{eqnarray}
        \ln|\mbox{\boldmath$D$}|
        =\int_0^\infty dx\,\ln x\,
        \frac{d^4f(x,0)}{dx^4}=\frac{\pi}{24}
        (ma)^3+O(m^2),\,\,\,a\gg a'.                  \label{incons}
        \end{eqnarray}
This implies the inconsistency of the regularised density matrix in view
of its infinitely growing off-diagonal elements.

The behaviour of the regularised density-matrix elements in close vicinity of its
diagonal is also unsatisfactory. For small $|\varphi-\varphi'|$
the dimensionally regularised decoherence factor in the limit of large mass
gives the result
                 \begin{equation}
                 \ln|\mbox{\boldmath$D$}(t|\varphi,\varphi')| \simeq
                 \frac{7}{64}m^3\bar{a}(a-a')^2\ ,
                 \label{wrongsign}
                 \end{equation}
also leading to the inconsistency of the density matrix.

The physical inconsistency of the dimensional regularisation is even more
transparent in applications to massless conformal invariant fields. On
Friedmann backgrounds, such fields decouple from the
gravitational variables, so one should expect the complete absence of
decoherence effects, $\mbox{\boldmath$D$}(t|\varphi,\varphi')=1$. This is,
however, not the case. Simple calculations show that for a massless conformally
coupled scalar field the basis function $v_n^*(t)$ and the frequency
function have the form:
                \begin{eqnarray}
                &&v_n^*(t) = \frac{1}{\cosh Ht}
                \left(\frac{1 + i \sinh Ht}
                {1 - i \sinh Ht}\right)^{\frac{n}{2}},    \label{conform-basis}\\
                &&\Omega_n=na^2
                + i a^2\sqrt{H^2a^2-1}.   \label{conform1}
                \end{eqnarray}
The corresponding decoherence factor is then given by the divergent sum,
        \begin{eqnarray}
        &&\ln|\mbox{\boldmath$D$}|
        =-\frac14\sum_{n=1}^\infty n^2\,
        \ln \left(A+\frac B{n^2}\right),                        \\
        &&A=\left(\frac{a^2+a'^2}{2a\,a'}\right)^2,\,\,
        B=\left(\frac{a^2\sqrt{H^2a^2-1}
        -a'^2\sqrt{H'^2a'^2-1}}{2a\,a'}\right)^2.
        \end{eqnarray}
Its dimensional regularisation gives the exact result
        \begin{eqnarray}
        \ln|\mbox{\boldmath$D$}(t|\varphi,\varphi')|
        =\frac\pi{12}\left(\frac BA\right)^{3/2}\ ,          \label{conform10}
        \end{eqnarray}
again demonstrating the wrong behaviour of the off-diagonal elements of
the density matrix in view of the obvious growth of the exponential
$(B/A)^{3/2}\to\infty,\,\, |\varphi-\varphi'|\to\infty$.
For a minimally coupled scalar field the situation is analogous to the conformal
case.

Thus, in the massless limit the regularised decoherence matrix turns out
to be as inconsistent as for massive fields. Moreover, for conformally
invariant fields it leads to an anti-intuitive conclusion of growing
quantum correlations between the environment and gravitational background
in spite of their actual decoupling.

\section{Conformal parametrisation of bosonic fields}
\hspace{\parindent}
In this section we discuss a possible solution to the problems
encountered with dimensional regularisation. It was remarked in the
Introduction that a redefinition of the environmental fields can
change the reduced density matrix. Moreover, as was remarked in
\cite{LL}, such a redefinition can be motivated by physical
considerations. In fact, it was already shown
for a specific model in \cite{Claus} that a redefinition can lead
both to a finite decoherence factor and to a consistent density matrix.
Here we shall demonstrate that such a viable redefinition is
not arbitrary and can lead to a unique result.

Let us consider the following redefinition of a bosonic scalar field,
        \begin{eqnarray}
        f(t)\rightarrow\tilde{f}(t)=a^\mu(t)\,f(t)          \label{6.1}
        \end{eqnarray}
with an arbitrary numerical parameter $\mu$. Under this redefinition the
kinetic term of the Lagrangian changes as
        \begin{eqnarray}
        L(f)=\frac12 a^3\left(\frac{df}{dt}\right)^2+...=
        \frac12 a^{3-2\mu}\left(\frac{d\tilde{f}}{dt}\right)^2+...\label{6.2}
        \end{eqnarray}
and the Wronskian operator and basis functions get replaced by
        \begin{eqnarray}
        &&{W}(d/dt)=a^3\frac d{dt}\rightarrow\tilde{W}(d/dt)=
        a^{3-2\mu}\frac d{dt},                                 \label{6.3}\\
        &&v_n(t)\rightarrow\tilde{v}_n(t)=a^\mu(t)\,v_n(t).    \label{6.4}
        \end{eqnarray}
The new frequency function then reads
        \begin{eqnarray}
        &&\tilde{\Omega}_n=-ia^{3-2\mu}\frac d{dt}\ln\tilde{v}_n^* \nonumber\\
        &&\qquad=
        a^{2-2\mu}\left[\sqrt{n^2+m^2a^2}
        +i\sinh Ht\left(1-\mu+\frac12\frac{m^2a^2}{n^2+m^2a^2}
        \right)\right]+O(1/m).
        \end{eqnarray}

For a particular choice of the parameter $\mu=1$ two important things
happen with this function -- the leading $n$ behaviour of $\tilde{\Omega}_n$
becomes independent of the macroscopic variable, and its imaginary part
decreases with growing $n$ as $1/n^2$,
        \begin{eqnarray}
        &&\tilde{\Omega}_n=\sqrt{n^2+m^2a^2}+b_n, \\
        &&{\rm Re}\,b_n=O(1/n),  \\
        &&{\rm Im}\,b_n=
        \sinh Ht\,\frac12\frac{m^2a^2}{n^2+m^2a^2}+O(1/m)=O(1/n^2).
        \label{finite}
        \end{eqnarray}
Substituting this behaviour into (\ref{4.4}), one finds
        \begin{eqnarray}
        \ln\tilde{\mbox{\boldmath$D$}}(t|\varphi,\varphi')\simeq\frac14
        \sum_{n=0}^\infty n^2\,\left(\,i\frac{{\rm Im}\,b_n'}n-
        i\frac{{\rm Im}\,b_n}n+O(1/n^4)\,\right)\ .
        \end{eqnarray}
This expression has at most logarithmic divergences which are imaginary
and, thus, affect only the phase of the density matrix. Moreover, since $b_n$
and $b_n'$ depend on $a$ and $a'$, respectively, these divergences
decompose into an additive sum of one-argument functions and,
therefore, can be cancelled by counterterms to the classical action $S$
(and $S'$) in (\ref{4.1}). The real part is simply convergent and
generates the finite decoherence amplitude. This result is formally
similar to the result for the decoherence factor in QED \cite{Claus}.

Let us show now that no inconsistencies arise with this redefined density
matrix. To begin with, let
us consider the far off-diagonal behaviour of the decoherence-factor
amplitude for the arbitrary parameter of reparametrisation $\mu$
 in the large-mass
limit. In this limit $\tilde{\Omega}_n\simeq a^{2-2\mu}\sqrt{n^2+m^2a^2}$,
and the decoherence factor can approximately be represented by the
integral (replacing the summation, $n=may$)
        \begin{eqnarray}
        \ln|\tilde{\mbox{\boldmath$D$}}(t|\varphi,\varphi')|\simeq\frac{(ma)^3}4
        \int_0^\infty dy\, y^2\ln\frac{4\delta^{2-2\mu}\sqrt{1+y^2}\,
        \sqrt{y^2+\delta^2}}{\left(\sqrt{1+y^2}
        +\delta^{2-2\mu}
        \sqrt{y^2+\delta^2}\right)^2},              \label{}
        \end{eqnarray}
where $\delta=a'/a$. This integral is convergent {\em only} for $\mu=1$, and
for $a\gg a'$ it yields the answer which shows the suppression of far
off-diagonal terms,
        \begin{eqnarray}
        |\tilde{\mbox{\boldmath$D$}}(t|\varphi,\varphi')|\simeq\exp\left[
        -\frac{(ma)^3}{24}\left(\pi-\frac83\right)
        +O(m^2)\right].                                     \label{6.5}
        \end{eqnarray}
For other values of $\mu$ this integral should be regulated to give the
final answer. Dimensional regularisation in the main changes the power of
$y$ in the integration measure from $y^2$ to $y^{2-\varepsilon}$. Then the
integration gives two different results for different ranges of $\mu$
        \begin{eqnarray}
        &&|\tilde{\mbox{\boldmath$D$}}(t|\varphi,\varphi')|\simeq\exp\left[\,
        -\frac{\pi}{24}(ma)^3\,\right],\,\,\,\mu>1,  \\
        &&|\tilde{\mbox{\boldmath$D$}}(t|\varphi,\varphi')|\simeq\exp\left[\,
        \frac{\pi}{24}(ma)^3\,\right],\,\,\,\mu<1\ .
        \end{eqnarray}
In the latter case this obviously coincides with the behaviour of the
inconsistent dimensionally regularised density
matrix (\ref{incons}) that was naively defined for $\mu=0$. Thus, although
a sensible result can be obtained for $\mu>1$, the value
$\mu=1$ is distinguished in yielding directly a finite result.
In spite of this, the results for $\mu=1$ and $\mu>1$ are qualitatively
the same.

In the vicinity of the diagonal the modified density matrix also acquires
a good quasi-Gaussian behaviour. Expansion of the expression given above
in powers of small $|a-a'|$ gives for large $m$ the dominant result
        \begin{equation}
        \ln|\tilde{\mbox{\boldmath$D$}}(t|\varphi,\varphi')| =
        -\frac{m^3\pi\bar{a}(a-a')^2}{64}.                  \label{finite1}
        \end{equation}

For massless conformally-invariant fields, the modified density matrix
justifies the expectations induced by the decoupling of the environment
from the gravitational background, cf. \cite{LL}.
 Its decoherence factor turns out to be
trivial. This can be seen by noting that, say,
in the case of massless conformally-invariant
scalar field, the basis and frequency functions have the form:
                \begin{eqnarray}
                &&\tilde{v}_n^*(t) =
                \left(\frac{1 + i \sinh Ht}{1 - i \sinh Ht}
                \right)^{\frac{n}{2}},               \label{conf-bas}\\
                &&\tilde{\Omega}_n =
                -i a \frac{d}{dt}
                \ln \tilde{v}_n^*(t) = n \ .            \label{conf1}
                \end{eqnarray}
Hence, $\tilde{\mbox{\boldmath$D$}}(t|\varphi,\varphi')\equiv 1$.
The same holds also for the electromagnetic field.

We conclude this section by briefly considering two massless conformally
non-invariant fields -- a minimally coupled scalar field and gravitons. What
they have in common are the basis- and frequency functions in their
respective conformal parametrisations:
                \begin{eqnarray}
                &&\tilde{v}_n^*(t) =
                \left(\frac{1 + i \sinh Ht}{1 - i \sinh Ht}
                \right)^{\frac{n}{2}}
                \left(\frac{n - i\sinh Ht}{n + 1}\right), \\
                &&\tilde{\Omega}_n = \frac{n(n^2-1)}{n^2-1+H^2a^2}
                - i \frac{H^2a^2\sqrt{H^2a^2-1}}{n^2-1+H^2a^2}.  \label{min1}
                \end{eqnarray}
They differ only by the range of quantum number $n$ ($2\leq n$ for
inhomogeneous scalar modes and $3\leq n$ for gravitons) and by the
degeneracies of the $n$-th eigenvalue of the Laplacian,
        \begin{eqnarray}
        &&{\rm dim} (n)_{\rm scal}=n^2\ , \\
        &&{\rm dim} (n)_{\rm grav}=2(n^2 - 4).
        \end{eqnarray}
The decoherence factor generated by both of these quantum fields thus has the
form of the following, obviously convergent, series:
        \begin{eqnarray}
        &&\ln|\tilde{\mbox{\boldmath$D$}}(t|\varphi,\varphi')| =-\frac14
        \sum_{n=2}^{\infty} {\rm dim} (n)\,\ln\left[\,1+
        \frac{\left(\sinh^2Ht-\sinh^2H't\right)^2}
        {4(n^2+\sinh^2Ht)(n^2+\sinh^2H't)}
        \right.                                            \nonumber\\
        &&\quad
        \left.+\frac{\left(\cosh^2Ht\,\sinh Ht\,(n^2+\sinh^2H't)-
        \cosh^2H't\,\sinh H't\,(n^2+\sinh^2Ht)\right)^2}{4n^2(n^2-1)^2
        (n^2+\sinh^2Ht)(n^2+\sinh^2H't)}\,\right]. \label{series}
        \end{eqnarray}
The calculation of the corresponding far off-diagonal elements is more
complicated here, but qualitatively it reproduces the behaviour
(6.11) for a massive scalar field with the mass parameter replaced by
the Hubble constant of the gravitational background,
        \begin{eqnarray}
        |\tilde{\mbox{\boldmath$D$}}(t|\varphi,\varphi')|\sim
        e^{-C(Ha)^3},\,\,\,a\gg a',\,\,\, C>0.
        \end{eqnarray}

To study the behaviour of matrix elements in the vicinity of diagonal of
density matrix it is enough to expand the expression (\ref{series})
up to the second order in the quantity $(H-H')$ and calculate the arising
series by using \cite{Grad-Ryz}. The exact result even in
this order of $(H-H')$ is too long to be fully presented here. Thus we
restrict ourselves to the limit of late time $t$ for this
expression,
        \begin{eqnarray}
        |\tilde{\mbox{\boldmath$D$}}(t|\varphi,\varphi')|\sim
        \exp\left(-\frac{\pi^2}{32} (H-H')^2 t^2 e^{4Ht}\right),\,\,\,
        Ht\gg 1,
        \end{eqnarray}
which clearly shows a rapid growth of decoherence during the inflationary
evolution.

\section{Conclusions}
\hspace{\parindent}
Let us recapitulate briefly the main results of the present paper.
We have calculated the reduced density matrix for the inflaton field
in a model of chaotic inflation by tracing out inhomogeneous
degrees of freedom corresponding to various bosonic fields. In the
original parametrisation of these fields, ultraviolet divergences arise in the
decoherence factor of the density matrix. The dimensional
regularisation of these divergences was shown to violate the consistency
of a reduced density matrix as a bounded operator, which is apparently related
to the fact that these divergences do not have the structure of one-field
counterterms usually used for ultraviolet renormalisation. A physically
motivated conformal redefinition of the environmental bosonic fields
leads, however, to well-defined finite results, which show that the Universe
acquires classical properties near the onset of inflation.

The seemingly mysterious finiteness property for the decoherence factor
in the conformal parametrisation has a natural explanation.
The off-diagonal elements of the density matrix, in the language of Feynman
diagrammes, are roughly proportional to the series of gravitational-matter
vertices with two quantum matter legs and the growing number of the
gravitational (background fields) ones. Indeed, when expanded in Taylor series
in $a-a'$ (and $\varphi-\varphi'$), the one-loop density matrix can be
represented as a quantum loop with the insertion of external legs
corresponding to the differentiations with respect to background fields,
these insertions forming the vertices in question. Such vertices generally
contain two derivatives which make the loop graph strongly
divergent. The conformal parametrisation of quantum matter fields on the
Friedmann background effectively removes the gravitational
variables from the kinetic term of the matter Lagrangian, thus actually
removing the derivatives from the corresponding vertex. This makes the
relevant one-loop Feynman diagrammes finite.

The obtained results show that the decoherence properties of the system
strongly depend on the parametrisation of the quantum fields that are traced
out. Although, intuitively, this sounds unsatisfactory -- the observable
correlations depend on the choice of variables about which we all the same
do not have any information -- this situation is very typical in quantum
theory. In fact, the density matrix is analogous to the S-matrix taken off
shell. As is well known, quantities such as
 the off-shell S-matrix, the off-shell effective
action,  strongly depend on the parametrisation of quantum fields, the
integration over which is performed when calculating these quantities.
A similar situation arises here: In contrast to the diagonal elements given by
the on-shell effective action, the off-diagonal elements are ambiguous. For
the bosonic fields that we considered here, this situation seems to have a
relatively satisfactory resolution -- there exists a distinguished parametrisation in
which divergences are absent. This parametrisation is distinguished for
several reasons. First, it reflects the conformal weight properties of the
field in question even in the case when this field is not conformally
invariant. Secondly, this parametrisation is attained by an overall
multiplication of the field by some power of the scale factor, which does
not spoil the locality of the field. As we shall show in a forthcoming
paper \cite{fermions}, the situation with fermions is  less pleasant.
Conformal parametrisation leaves the answers divergent, while the
additional reparametrisation that can make them ultraviolet-finite is
nonlocal -- it has the form of Bogoliubov transformations rotating the
different harmonics of the field differently, thus destroying the local
nature of the original field.

The results obtained have strong implications for the
quantum-to-classical transition within the theory of the quantum origin of
the inflationary Universe. They can be especially important in the quantum
cosmology of the chaotic inflationary model with big negative nonminimal coupling
$\xi$ of the inflaton field to curvature \cite{qsi,tvsnb,qcr1,efeq}. This model
was shown to generate a probability peak in the
distribution function of the inflaton field $\varphi$ -- the diagonal element of
the density matrix -- at the GUT energy scale for the inflationary Hubble constant
$H(\varphi)\sim m_{\rm P}\sqrt\lambda/|\xi|\sim 10^{-5}m_{\rm P}$ ($\lambda$ is
a coupling
constant in the quartic term of the inflaton potential, and the value of the
ratio $\sqrt\lambda/|\xi|\sim 10^{-5}\sim \Delta T/T$ follows in this model from
the COBE normalisation of the magnitude of the microwave background
anisotropy). This probability peak has at the onset of inflation a very narrow quantum
width for the Hubble constant, $\Delta H/H\sim\sqrt\lambda/|\xi|$. Now we can show
that the width of the density matrix in the off-diagonal direction at the
initial moment of time $t=0$ has the same magnitude. This follows from
(\ref{4.2}).

First, apart from the decoherence factor
$\tilde{\mbox{\boldmath$D$}}(0|\varphi,\varphi')$ the only strongly suppressing
factor in this equation is, cf. (\ref{effect}),
        \begin{eqnarray}
        \exp(-\mbox{\boldmath$\SGamma$}_{\rm 1-loop}/2
        -\mbox{\boldmath$\SGamma$}'_{\rm 1-loop}/2)
        \sim (HH')^{-Z/2},
        \end{eqnarray}
with a very big parameter of anomalous scaling in this model, $Z\sim
|\xi|^2/\lambda$. It gives a power-like suppression of the far off-diagonal
terms, and the quantum width in the off-diagonal direction, in the
vicinity of the diagonal,
        \begin{eqnarray}
        (HH')^{-Z/2}\sim H^{-Z}\,e^{-(H-H')^2/\Delta H^2},
        \end{eqnarray}
$\Delta H\sim H\sqrt\lambda/|\xi|$, coincides with the width of the wave
packet.

As far as the decoherence factor $\tilde{\mbox{\boldmath$D$}}$ is concerned,
at the start of inflation it is basically inefficient,
$\tilde{\mbox{\boldmath$D$}}(0|\varphi,\varphi')=1$, for a very simple
reason. Large $|\xi|$ generate in this model by the Higgs mechanism very big
values of masses for all quantum fields directly coupled to inflaton. The
order of magnitude of their mass parameters is such that the ratio
$m^2/H^2\sim |\xi|$ is very big and $\varphi$-independent. This means that
for big masses at $t=0$ the dominant argument of the decoherence factor
$ma(0)=m/H$ is $\varphi$-independent and
$\tilde{\mbox{\boldmath$D$}}(0|\varphi,\varphi')=1$. Thus, the Universe is
essentially quantum at the ``start of inflation''. But due to decoherence
effects it rapidly becomes classical: the quantum width of the
distribution function -- the probability -- stays basically the same,
while the off-diagonal width rapidly decreases. From (\ref{finite1}) it
immediately follows that this width for late time is
$\Delta H\sim \exp(-3Ht/2)/(t|\xi|^{3/4})$.

In our paper we have restricted ourselves to decoherence within one
semiclassical branch of the wave function. There is, however, also the
possibility that different branches might interfere. For the examples discussed
in our paper, this would be the case for the no-boundary state that is
a superposition of two complex conjugate semiclassical branches. Decoherence is,
however, also effective in suppressing interferences between such branches as long
as decoherence within one branch holds \cite{Claus}. At the technical level,
this follows from the expression (\ref{4.4}) which for the interbranch case
has in the denominator $\Omega_n+\Omega_n'$ rather than $\Omega_n+\Omega_n'^*$.
This means that the imaginary parts of the frequency functions add up instead of
partially cancelling one another. Therefore, the amplitude of the interbranch
decoherence factor is smaller than that of one branch. In quantum mechanics,
a nice analogy is provided by the case of chiral molecules
\cite{decoherence1} where the superposition between the left-handed and
the right-handed version of the corresponding molecule is
being suppressed after the localisation has been taken place by continuous
measurement with, for example, light in each version.

We have here not included a discussion of the influence of fermionic fields,
since this will need a different formalism and reveal novel aspects.
This case will be discussed in a separate paper \cite{fermions}
that, together with the present paper, should present a complete
picture of decoherence in one-loop quantum cosmology at the onset of
inflation.

\section*{Acknowledgements}
\hspace{\parindent}
We are grateful to A.A. Starobinsky for useful discussions. A.K.
also thanks V.L. Chernyak for a useful discussion.
The work of A.B. was partially supported by RFBR under the grant No
96-02-16287. The work of A.K. was partially supported by RFBR under
the grant No 96-02-16220 and under the grant for support of leading
scientific schools No 96-15-96458. A.B. and A.K. kindly acknowledge
financial support by the DFG grants 436 RUS 113/333/4 during their
visit to the University of Freiburg in autumn~1998. The work of A.K.
was also supported by the CARIPLO scientific foundation.


\begin{thebibliography}{99}
\bibitem{no-boundary}J.B. Hartle and S.W. Hawking,
        Phys. Rev. D 28 (1983) 2960;
        S.W. Hawking, Nucl. Phys. B 239 (1984) 257.
\bibitem{tun}A.D. Linde, JETP 60 (1984) 211, Lett. Nuovo Cim. 39 (1984)
        401; V.A. Rubakov, Phys. Lett. B 148 (1984) 280; A. Vilenkin,
        Phys. Lett. B 117 (1982) 25, Phys. Rev. D 30 (1984) 549;
        Ya. B. Zeldovich and A.A. Starobinsky,
        Sov. Astron. Lett. 10 (1984) 135.
\bibitem{CZ} H.D. Conradi and H.D. Zeh, Phys. Lett. A 154 (1991) 321;
            H.D. Conradi, Phys. Rev. D 46 (1992) 612.
\bibitem{qsi}A.O. Barvinsky and A.Yu. Kamenshchik,
              Phys. Lett. B 332 (1994) 270.
\bibitem{qcr1}A.O. Barvinsky, A.Yu. Kamenshchik and I.V. Mishakov,
        Nucl. Phys. B 491 (1997) 387.
\bibitem{Zeh} H.D. Zeh, Found. Phys. 1 (1970) 69; in: Foundations
        of Quantum Mechanics, ed. by B.~d'Espagnat (Academic Press,
        New York, 1971); Found. Phys. 3 (1973) 109; O. K\"ubler
        and H.D. Zeh, Ann. Phys. (N.Y.) 76 (1973) 405.
\bibitem{decoherence1} D. Giulini, E. Joos, C. Kiefer, J. Kupsch,
      I.-O. Stamatescu and H.D. Zeh, Decoherence and the Appearance of
     a Classical World in Quantum Theory (Springer, Berlin, 1996).
\bibitem{decoherence2} C. Kiefer and E. Joos, in: Quantum Future,
    ed. by P. Blanchard and A. Jadczyk (Springer, Berlin, 1998).
\bibitem{Haroche} M. Brune, E. Hagley, J. Dreyer, X. Ma\^{\i}tre, A. Maali,
     C. Wunderlich, J.M. Raimond and S. Haroche, Phys. Rev. Lett.
     77 (1996) 4887.
\bibitem{Joos} E. Joos, Phys. Lett. A 116 (1986) 6.
\bibitem{Zeh2} H.D. Zeh, Phys. Lett. A 116 (1986) 9.
\bibitem{Kiefer1} C. Kiefer, Class. Quantum Grav. 4 (1987) 1369.
\bibitem{Polarski} C. Kiefer, D. Polarski and A.A. Starobinsky,
    Int. J. Mod. Phys. D 7 (1998) 455; C. Kiefer and D. Polarski,
    Ann. Phys. (Leipzig) 7 (1998) 137; C. Kiefer, J. Lesgourgues,
    D. Polarski and A.A. Starobinsky, Class. Quantum Grav. 15 (1998) L67.
\bibitem{Paz}J.P. Paz and S. Sinha, Phys. Rev. D 45 (1992) 2823.
\bibitem{Claus}C.Kiefer, Phys. Rev. D 46 (1992) 1658.
\bibitem{Okamura}T. Okamura, Prog. Theor. Phys. 95 (1996) 95.
\bibitem{LL} R. Laflamme and J. Louko, Phys. Rev. D 43 (1991) 3317.
\bibitem{norm}A.O. Barvinsky and A.Yu. Kamenshchik, Class. Quantum Grav. 7
        (1990) L181.
\bibitem{tvsnb}A.O. Barvinsky and A.Yu. Kamenshchik, Int. J. Mod. Phys.
        D 5 (1996) 825.
\bibitem{norm-napoli}
G. Esposito, A.Yu. Kamenshchik and G. Miele,
Phys. Rev. D 56 (1997) 1328.
\bibitem{fermions} A.O. Barvinsky, A.Yu. Kamenshchik and C. Kiefer,
Effective action and decoherence for fermions in quantum cosmology,
in preparation.
\bibitem{Kiefer2} C. Kiefer, in: Canonical Gravity: From Classical
   to Quantum, ed. by J. Ehlers and H. Friedrich (Springer, Berlin, 1994).
\bibitem{Ringsberg} C. Kiefer, in: Time, Temporality, Now, ed. by
  H. Atmanspacher and E. Ruhnau (Springer, Berlin, 1997).
\bibitem{Laf}J.J. Halliwell and S.W.Hawking, Phys. Rev. D 31 (1985) 1777;
        R. Laflamme, Phys. Lett. B 198 (1987) 156.
\bibitem{VilVach}T. Vachaspati and A. Vilenkin, Phys. Rev. D 37 (1988) 898.
\bibitem{tunnel}A.O.Barvinsky and A.Yu.Kamenshchik, Phys. Rev.
        D 50 (1994) 5093.
\bibitem{BK}A.O. Barvinsky and V. Krykhtin, Class. Quantum Grav. 10 (1993)
           1957; A.O. Barvinsky, Class. Quantum Grav. 10 (1993) 1985.
\bibitem{annals}A.O. Barvinsky, A.Yu. Kamenshchik and I.P. Karmazin, Ann. Phys.
        (N.Y.) 219 (1992) 201.
\bibitem{BarvU}A.O. Barvinsky, Phys. Rep. 230 (1993) 237.
\bibitem{Allen}B. Allen, Phys. Rev. D 32 (1985) 3136.
\bibitem{asymp}
F.W.J. Olwer, Introduction to asymptotics and special functions
(Academic Press, New York-London, 1974);
R.C. Thorne, Phil. Trans. Roy. Soc. London 249 (1957) 597.
\bibitem{Grad-Ryz}
I.S. Gradstein and I.M. Ryzhik, Table of integrals, series, and
products, (Academic Press, New York, 1980).
\bibitem{efeq}A.O. Barvinsky and A.Yu. Kamenshchik, Nucl. Phys. B 532 (1998)
        339.
\end{thebibliography}
\end{document}